\documentclass[%
 reprint,
 amsmath,amssymb,
 aps,
]{revtex4-1}

\usepackage{graphicx}
\usepackage{dcolumn}
\usepackage{bm}

\usepackage{physics}
\usepackage{graphicx}
\usepackage[export]{adjustbox}
\usepackage{subfig}
\usepackage{float}

\usepackage[size=scriptsize]{caption}

\begin{document}

\preprint{APS/123-QED}

\title{Engineering and Suppression of Decoherence  in Two Qubit Systems}

\author{Govind Unnikrishnan}
\affiliation{%
 Indian Institute of Science Education and Research (IISER), Pune, Maharashtra, India\\
}%

\date{\today}

\begin{abstract}
In this work, two experimentally feasible methods of decoherence engineering-one based on the application of stochastic classical kicks and the other based on temporally randomized pulse sequences are combined. A different coupling interaction is proposed, which leads to amplitude damping as compared to existing methods which model phase damping, utilizing the $ zz$ coupling interaction. The decoherence process on combining the stochastic kick method and the randomized pulse sequence method and the effectiveness of dynamical decoupling under these coupling interactions are analyzed. Finally,a counter intuitive result where decoherence is suppressed in the presence of two noise sources under certain resonant conditions is presented. 
\end{abstract}

\pacs{Valid PACS appear here}
\maketitle


\section{\label{sec:level1}Introduction}
In quantum information processing, we often make use of coherent superpositions of quantum states to achieve an advantage over classical analogues. In order to preserve coherence, one may try to isolate the system from external influences, but this will come at the expense of losing the ability to manipulate the system. There are alternate methods such as quantum error correction, dynamical decoupling and encoding in decoherence free subspaces which help protect the system coherence\cite{qerr,dd}. However, in all realistic systems, there is an inevitable and irreversible loss of information or energy from the system due to interactions with the environment. This loss is termed decoherence. We may classify decoherence into two types-phase decoherence (loss of information) and spin flip decoherence (loss of energy), using which more complex decoherence processes may be constructed \cite{jones}. In order to suppress decoherence, it is crucial to understand the underlying mechanism and dynamics of decoherence processes. This is the motivation for engineered decoherence, which serves as a testbed for understanding the physics behind natural decoherence processes in a controlled manner, thus enabling the improvement of existing decoherence suppression strategies and facilitating the creation of new, improved suppression methods.     

As decoherence is a challenging problem cutting across various qubit realizations ranging from NV centers in diamond to trapped ion qubits, a very general model independent of the physical system chosen to implement the qubit is desirable. Zurek \cite{zurek} first proposed such a model in which he considered an N-qubit system in which each qubit interacted with another via the $\sigma _z \sigma _z $ interaction. However, in this model one requires N to be very large in order to accurately model the damping of system coherence, rendering it experimentally unfeasible. Building on Zurek's model, Cory's group demonstrated an experimentally feasible method \cite{Cory} to effectively model phase damping in a system with finite number of qubits. By applying random amplitude kicks to the environmental qubit and averaging over many realizations, the need for large number of environmental qubits is made unnecessary.    

A different approach to model phase damping in finite dimensional systems was proposed by Kondo et al \cite{kondo}. In this model, only pairs of $ \pi $ pulses are applied to the environmental qubit but the time period between successive pulses are randomized to obviate the necessity of a large number of environmental qubits. On averaging over different realizations of the delay between immediate $ \pi $ pulses, one realizes the phase damping model. It is worth pointing out that both these models are based on Zurek's \cite{zurek} initial proposal and they differ mainly in the method of introducing randomness into the system to eliminate the need for a large system size. However, Kondo's model is much simpler to solve analytically, even in the presence of control fields \cite{kondo}.     

In this paper, we first see how the introduction of $ x x $ coupling in the place of $ z z $ coupling in Zurek's model will lead to spin flip decoherence, and further show that Cory's experimentally feasible method is equally effective in this case. We then analyze the effect of a dynamical decoupling sequence on the two fundamental types of decherence-phase decoherence and spin flip decoherence. On combining Cory's method \cite{Cory} of random amplitude kicks and Kondo's method \cite{kondo} of using $ \pi $ pulses at random time intervals, although one expects the decoherence rate to be enhanced, at certain resonant frequency regimes, quite counter intuitively, we observe a suppression of decoherence in the dephasing model. Furthermore, it is shown that this suppression of decoherence outperforms the standard dynamical decoupling sequence (a series of equidistant $ \pi $ pulses) at certain low frequency ranges.    

This paper is organized as follows. In section II, Cory's model \cite{Cory} of phase damping is introduced, which in turn is grounded on Zurek's model \cite{zurek}. Section III gives a brief overview of the temporal randomization technique employed by Kondo et al \cite{kondo} to model dephasing. We introduce the $ x x $ interaction in Zurek's model in section IV and obtain an analytical solution which shows that the $ x x $ interaction leads to spin flip decoherence. In section V, results of numerical simulations obtained on combining Cory's model and Kondo's model, for both phase damping and spin flip decoherence, are presented. In particular, we will see that decoherence gets suppressed in certain resonant frequency regimes on combining these two models. We conclude with a summary and discussion in section VI.        

\section{\label{sec:level2}Random Kick Model}
Zurek considered $ n $ two level systems interacting via the $ z z $ interaction in order to model dephasing \cite{zurek}. One takes $ n=1 $ as the system of interest and consider the interaction between this qubit and the remaining two level systems.  

\begin{equation}
H_{int}=\sum_{k=2}^{n} J_{1k} \sigma_z^1 \sigma_z^k .
\label{zmodel}
\end{equation}  

We evolve the combined system-environment density matrix for a time $ t $ and then trace out the environmental degrees of freedom to obtain the system density matrix.
\[
\rho^{S}  (t)= Tr_{E} \{ \rho^{SE} (t) \}                      
\]

\[
\rho^{S}  (t)= \begin{pmatrix} 
\rho_{00}^S (t)&\rho_{01}^S (t)\\
\rho_{10}^S (t)&\rho_{11}^S (t)\\
\end{pmatrix}.
\]

The diagonal terms, $\rho_{00} $ and $ \rho_{11} $, remain unchanged under $ z z $ interaction while the off diagonals, which characterize the system coherence, evolve. Assuming the initial environmental qubits to be in an arbitrary states $ \ket{\phi}_k=\alpha_k \ket{0}_k+\beta_k \ket{1}_k $ and the system to be in the initial pure state $ \ket{\psi}=a \ket{0}+b \ket{1} $, the system coherence at time $ t $ is calculated as 
\[
\rho_{01}^S(t)=ab*z(t)
\]   
with 
\begin{equation}
z(t)=\prod_{k=2}^{n} \abs{\alpha_k}^2 \exp(-2iJ_{1k} t)+ \abs{\beta_k}^2 \exp(2iJ_{1k} t).
\label{zt}
\end{equation}

It has been shown \cite{zurek} that $ z(t) $ will return arbitrarily close to it's initial value unless $ n\rightarrow \infty $, an unfeasible situation in physical realizations.

Cory's group proposed a simple, solvable model \cite{Cory} composed of one environmental qubit (E) and one system qubit(S).
\begin{equation}
H_{0}=  \pi  (\nu_S \sigma_{z}^{S} + \nu_E \sigma_{z}^{E}+ \frac{\Omega}{2} \sigma_{z}^{S} \sigma_{z}^{E}) 
\label{cory_hamilt}
\end{equation}  
where $ \nu_S$ and $ \nu_E $ are chemical shifts and $ \frac{\Omega}{2} $ is the coupling strength.

In order to compensate for the limited number of environmental qubits, random amplitude kicks are applied to the environmental qubit, which are of the form
\[
K^E_m=\exp(-i\epsilon_m \sigma_y^E)
\] 
where $ \epsilon_m $ is randomly chosen from the range  ($ -\alpha, \alpha$). For a total evolution time  T , these kicks are applied at every $ T/n $ intervals. We may define the kick rate $ \Gamma=\frac{n}{T} $. The evolution operator is thus
\[
U_{total}(T)=K_n U(T/n) K_{n-1} U(T/n)...K_{1} U(T/n)
\]
where U(t)=$ \exp{-iH_0t} $.

The system density matrix at time T is obtained by averaging over all realizations of $ \epsilon_m $ and tracing out the environmental degree of freedom
\begin{equation}
\overline{\rho^S(T)}=\int_{-\alpha}^{\alpha} \frac{d\epsilon_n}{2\alpha}...\int_{-\alpha}^{\alpha} \frac{d\epsilon_1}{2\alpha} Tr_{E} \{ U_{tot}(T) \rho^{SE} (0) U_{tot} ^{\dagger}(T)\}.
\end{equation}

On simplifying this further by considering an initial factorisable state and expanding in the eigenbasis of $ \sigma_z $, one obtains
\begin{equation}
\overline{\rho^S(T)}=\sum_{i,j=0,1}{} \rho_{ij}^{S}(0)f_{ij}(T,n) \ket{i} \bra{j} 	.
\end{equation}  
The decoherence factor, $ f_{ij} (T,n)$ is calculated as
\begin{equation}
f_{ij}(T,n)=\int_{-\alpha}^{\alpha} \frac{d\epsilon_n}{2\alpha}...\int_{-\alpha}^{\alpha} \frac{d\epsilon_1}{2\alpha} Tr_{E} \{ (A^{E}_{i})_n \rho^{E} (0) (A^{E}_{j})_n\}
\end{equation}
where
\begin{equation}
(A^{E}_{j})_n= K^{E}_{n}V^{E}_{j}...K^{E}_{2}V^{E}_{j}K^{E}_{1}V^{E}_{j}.  
\end{equation}
The operator $ V^{E}_j $ is obtained after tracing out the system qubit from $ U(T/n) $ i.e. $ V^{E}_{j}= _S\bra{j}U(T/n)\ket{j}_S $
It is clear that $ f_{jj}=1 $. Hence, the final system density matrix is
\[
\overline{\rho^S(T)}=\begin{pmatrix}
\rho_{00}^{S}(0)&f_{01}(n,T)\rho_{01}^{S}(0)\\
f_{01}^{*}(n,T)\rho_{10}^{S}(0)&\rho_{11}^{S}(0) \\
\end{pmatrix}.
\]

The system coherence is quantified by the off diagonal term $ f_{01} $(n,T), which goes to zero for appropriate limits of $ \alpha $ and $ \Gamma $ (the kick rate), thus describing phase damping.

\section{Kondo's method}
\begin{figure}[H]
		\caption{Schematic representation of Kondo's method: First $ \pi $ pulse is applied at time $ t_1 $ and second $ \pi $ pulse at time $ t_1+\delta $}
		\includegraphics*[width=0.5\textwidth]{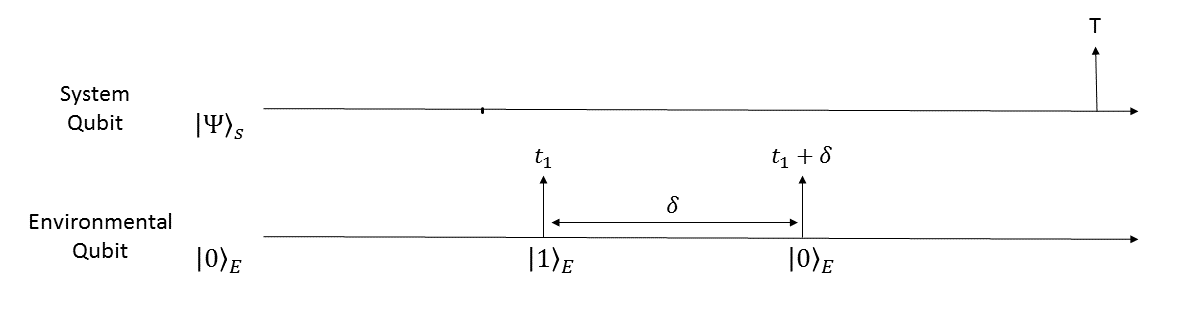}
		\label{fig1}
	\end{figure}

	Instead of the random kicks used by Cory, Kondo et al suggested sequence of temporally randomized $ \pi $ pulses to preempt the use of a large number of environmental qubits \cite{kondo}. Here also, an interaction of the kind in equation (\ref{zmodel}) is assumed with $ n=2 $. We suppose that the initial state of the environmental qubit is $ \ket{0} $ and that it is flipped by a $ \pi $ pulse to $ \ket{1} $ at a time $ t_1 $. At a time $ t_1+\delta $, it is flipped back into it's initial state $ \ket{0} $ and we make our observation of the system qubit at a time T. It is schematically shown in figure~\ref{fig1}. 
	Here, we have two situations
	\begin{equation}
	H_{int} \ket{\Psi} \ket{0}=J_{12}\sigma_z^S\ket{\Psi}\ket{0} 
	\label{op1}
	\end{equation}
		
	\begin{equation}
	H_{int} \ket{\Psi} \ket{1}=-J_{12}\sigma_z^S\ket{\Psi}\ket{1}		\label{op2}.
	\end{equation}
	
	This means that two different operators (\ref{op1}) and (\ref{op2}) act on the system qubit, conditional on the state of the environmental qubit. To evaluate the state of the system qubit at time T, we apply the unitary
	\begin{equation}
	e^{-iJ_{12} \sigma_{z} (T-t_{1}- \delta)} e^{iJ_{12}\sigma_{z} \delta} e^{-iJ_{12} \sigma_{z} t_{1}}= S(2J_{12} \delta) S(-J_{12} T)
	\end{equation}
	where 	$ S(\theta )=e^{i \theta \sigma_z} $ . Assuming that $ \delta $ takes on random values from the range $ 0 \leqslant 2J_{12} \delta \leqslant 2\pi$ from time 0 to T, the resultant system density matrix at time T is given by:
	
	\begin{equation}
	\overline{\rho^{S} (T)}= \frac{1}{2 \pi} \int_{0}^{2 \pi} d \theta S(\theta) \rho^S(0) S(\theta) ^{\dagger}
	\end{equation}
	where we have taken $ \theta=2J_{12} \delta $. This simplifies to 
	\[
	\overline{ \rho^{S}  (T)} = \begin{pmatrix} 
	\rho_{00}^S (0)& \expval{e^{i \theta}} \rho_{01}^S (0)\\
	\expval{e^{-i \theta}} \rho_{10}^S (0)&\rho_{11}^S (0)\\
	\end{pmatrix}.
	\]
	\\For $ \theta \in [0, 2 \pi],  \expval{e^{i \theta}}=0 $. Hence,
	
	\begin{equation}
	\overline{ \rho^{S}  (T)}= \begin{pmatrix} 
	\rho_{00}^S (0)& 0\\
	0&\rho_{11}^S (0)\\
	\end{pmatrix}.
	\label{zDM}
	\end{equation} 
	which indicates complete dephasing.	
	
\section{Spin Flip Decoherence}
In this section, we introduce the $ xx $	interaction into Zurek's model \cite{zurek} and show how this leads to spin flip decoherence. We look at n two level systems with a coupling Hamiltonian of the form
\begin{equation}
H_{SE}=\sum_{k=2}^{n} J_{1k} \sigma_x^1 \sigma_x^k.
\end{equation}
With the corresponding unitary operator
\begin{equation}
U_{SE}(t)=e^{-iH_{SE}t}=e^{-i \sum_{k=2}^{n} J_{1k} \sigma^1_x \sigma^k_x t}.
\end{equation}

Now, we consider a factorisable initial state and express both the system qubit (k=1) and the environmental qubits (k$ \geqslant $2) in the eigenbasis of $ \sigma_x $. Dropping the subscript k=1 for the system qubit,

\begin{equation}
\ket{\Psi(0)}_{SE}=(a' \ket{+}+b' \ket{-}) \otimes \prod_{k=2}^{n} (\alpha'_k \ket{+}_k+\beta'_k \ket{-}_k ).
\end{equation} 	
The combined state at a time $ t $ is given by
\begin{align}
\ket{\Psi(t)}_{SE} &= U_{SE}(t)\ket{\Psi(0)}_{SE} \nonumber \\
&=  a' \ket{+} \otimes \prod_{k=2}^{n} e^{-iJ_{1k} \sigma_x t} (\alpha'_k \ket{+}_k+\beta'_k \ket{-}_k )\nonumber \\
&+ b' \ket{-} \otimes \prod_{k=2}^{n} e^{iJ_{1k} \sigma_x t} (\alpha'_k \ket{+}_k+\beta'_k \ket{-}_k ) \nonumber\\
&=  a' \ket{+} \otimes \prod_{k=2}^{n} e^{-iJ_{1k}t} \alpha'_k \ket{+}_k+ e^{iJ_{1k}t} \beta'_k \ket{-}_k \nonumber\\
&+ b' \ket{-} \otimes \prod_{k=2}^{n} e^{iJ_{1k} t} \alpha'_k \ket{+}_k+ e^{-iJ_{1k} t} \beta'_k \ket{-}_k )
\label{se}.
\end{align} 

Now, we trace out all the environmental qubits in order to get the system density matrix in the $ \{ \ket{+}, \ket{-} \} $ basis	
\begin{align*}
\rho^S(t)&= Tr_{E} \{\rho^{SE} (T)\} \\
&= Tr_{E} \{\ket{\Psi(t)} _{SE SE} \bra{\Psi (t)} \}.   
\end{align*}

Plugging in equation (\ref{se}) and simplifying using the constraints $ \alpha_{k}^{'2}+\beta_k^{'2}=1 $ and $ a'^2+b'^2=1 $, we obtain
\begin{align} 
\rho^{S} (t)_{\{\ket{+},\ket{-}\}} & =a'^2 \ket{+} \bra{+}  + a'b'^* z(t) \ket{+} \bra{-} \nonumber \\
&+ a'^*b' z^*(t) \ket{-} \bra{+} + b'^2 \ket{-} \bra{-} 
\end{align}
where $ z(t) $ is given by (\ref{zt}) with $ \alpha_k , \beta_k $ replaced by their primed versions. This is the density matrix in the \{$ \ket{+}, \ket{-} $\} basis. For converting to the computational basis, we apply the Hadamard transform
\begin{equation}
\rho^{S} (t)_{\{\ket{0},\ket{1}\}}=H \rho^{S} (t)_{\{\ket{+},\ket{-}\}} H^{\dagger}
\end{equation}	
with
\[
H=\frac{1}{\sqrt{2}}\begin{pmatrix}
1 & 1 \\
1 & -1
\end{pmatrix}.
\]
On solving, we get
\begin{equation}
\rho^{S} (t)_{\{\ket{0},\ket{1}\}}= \frac{1}{2}\begin{pmatrix}
1+(\omega+\omega^*)           & a'^2-b'^2-(\omega-\omega^*)\\
a'^2-b'^2+(\omega-\omega^*) & 1-(\omega+\omega^*)
\end{pmatrix}.
\label{oscill}
\end{equation}

where $ \omega=a'b'^*z(t) $. In the limit $ n \rightarrow \infty $, we have $ \omega \rightarrow 0 $.
\begin{equation}
\rho^{S} (t)_{\{\ket{0},\ket{1}\}}= \frac{1}{2}\begin{pmatrix}
1           & a'^2-b'^2\\
a'^2-b'^2 & 1
\end{pmatrix}.
\label{xmodel}
\end{equation}
By considering the case $ n=2 $, we can clearly see that the term $ (\omega+\omega^*) $ causes oscillations in the population levels quantified by the diagonal elements of the density matrix. For large $ n $, we see that the populations of $ \ket{0} $ and $ \ket{1} $ become equal. It is worth noting that although the physical situations are very different, mathematically, the form of state (\ref{xmodel}) is related by a simple Hadamard transform to the state (\ref{zDM}). In the following sections, we will show that this model can be realized with finite resources using Cory's method. We will also incorporate Kondo's method into this model and analyze the effect of dynamical decoupling on this type of decoherence.  

\section{Combining Random Kicks and Temporally Randomized $ \pi $ Pulses}
	 In this section,  results of numerical simulations are presented, which reveal the dynamics of the system coherence in the case of $ z z $ interaction and the populations in the case of    $ x x $ interaction.In simulations, one assumes typical values for the parameters in (\ref{cory_hamilt}) viz. $  \frac{\Omega}{2}=150$ Hz, $\alpha= 0.11(\frac{\pi}{2}) $. For both interactions, analysis is done for three cases:
	 \begin{enumerate}
	 	\item Under the application of Cory's random amplitude kicks
	 	\item Under the application of Cory's random kicks and a dynamical decoupling (DD) sequence (a series of equidistant $ \pi $ pulses) to suppress decoherence.
	 	\item Under the combined application of Cory's random kicks and Kondo's temporally randomized  $ \pi $ pulse sequence
	 \end{enumerate}
	 We show that in case (3) above, although the decoherence rate is enhanced as expected at most kick rates, there is a suppression of decoherence  when the kick rate is close to the system-environment coupling, given by $ \frac{\Omega}{2} $. Furthermore, it can be shown that this suppression outperforms the dynamical decoupling sequence composed of a series of equidistant $ \pi $ pulses for low frequencies of the decoupling pulses. Another interesting fact is that when $ \frac{\Omega}{2\Gamma}=p $, where $ p $ is integer, there will be no decoherence on applying only Cory's kicks. 
	 \subsection{The $ z z $ interaction}
	 \label{subsec:zz interaction}
\begin{figure}[]
	 	\subfloat[]{
	 		\includegraphics[scale=0.4, width=0.5\linewidth]{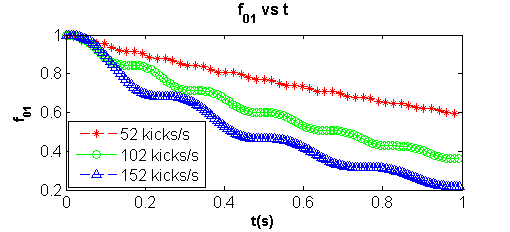}}
	 	\subfloat[]{
	 		\includegraphics[scale=1, width=0.5\linewidth]{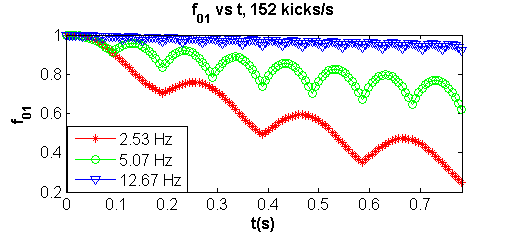}}
	 	\caption{Time evolution of the decoherence factor $ f_{01} (t)$ for (a) different kicks rates. (b) different decoupling frequencies. Coupling between qubits, $ \Omega/2 $=150 Hz.} 
	 	\label{fig2}		
	 \end{figure}

	 For convenience, we assume that the system starts from the initial state $ \rho^S(0)=\frac{1}{2}(I+\sigma_x) $ and the environmental qubit is in the thermal equilibrium state $ \rho^E(0)=\frac{1}{2}(I+\sigma_z) $. A larger kick rate leads to faster decoherence as shown in figure~\ref{fig2} (a) . However, Cory and group have shown that in the limit of very large kick rates, the system gets decoupled from the environment and decoherence is suppressed \cite{Cory}.						
	 
\begin{figure}[]
	 	\subfloat[]{
	 		\includegraphics[scale=0.4, width=0.5\linewidth]{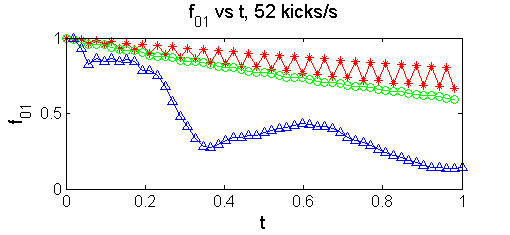}}
	 	\subfloat[]{
	 		\includegraphics[scale=0.4, width=0.5\linewidth]{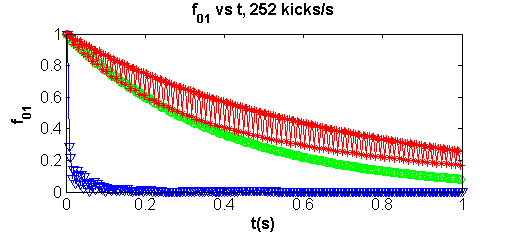}}
	 	\caption{Time evolution of the decoherence factor $ f_{01} (t)$ for (a) $ \Gamma $=52 kicks/s. Stars indicate the case where periodic $ \pi $ pulses are applied at 13 Hz, circles indicate the case where only Cory's kicks are applied and triangles show Cory's kicks and randomized $ \pi $ pulses.(b) For $ \Gamma $=252 kicks/s. Stars indicate the case where periodic $ \pi $ pulses are applied at 42 Hz, circles indicate the case where only Cory's kicks are applied and triangles show Cory's kicks and randomized $ \pi $ pulses. Coupling between qubits, $ \Omega/2 $=150 Hz.  } 
	 	\label{fig3}		
	 \end{figure}

\begin{figure}[]
	 		\subfloat[ ]{
	 			\includegraphics[width=0.5\linewidth]{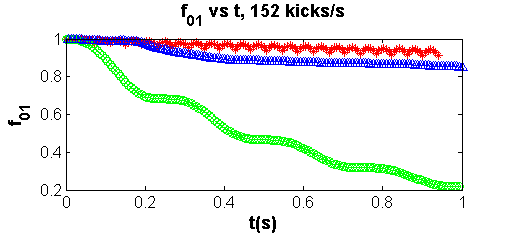}}
	 		\subfloat[]{\includegraphics[width=0.5\linewidth]{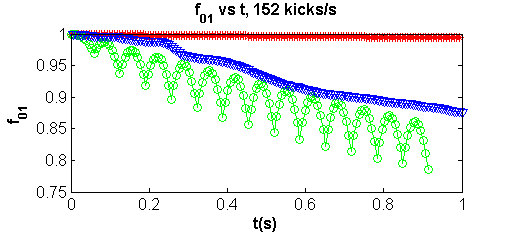}}
	 		\caption{Time evolution of the decoherence factor $ f_{01} (t) $ for $ \Gamma\approx \Omega/2 \equiv$152 kicks/s (a) Stars indicate the case where periodic $ \pi $ pulses are applied at 12.67 Hz, circles indicate the case where only Cory's kicks are applied and triangles show Cory's kicks and randomized $ \pi $ pulses. (b) Stars indicate the case where periodic $ \pi $ pulse frequency is 76 Hz, circles indicate the case where it is 7.6 Hz and triangles show Cory's kicks and randomized $ \pi $ pulses, One can see that the coherence is better preserved when random $ \pi $ pulses are applied than when periodic $ \pi $ pulses are applied at 7.6 Hz.} 
	 		\label{fig4}		
	 	\end{figure}

On incorporating a randomized $ \pi $ pulse sequence to Cory's model, we observe that the system coherence drops very quickly compared to the decay under Cory's kicks alone, at most kicks rates as shown in figure ~\ref{fig3}. However, when the kick rate is close to the system-environment coupling, we observe a suppression of decoherence, as show in in figure ~\ref{fig4}(a), when compared to only applying Cory's kicks. Furthermore, we see that this suppression due to random $ \pi $ pulses can outperform low frequency dynamical decoupling sequences when the kick rate is close to the resonant frequency of $ \Omega/2 $. This is shown in figure ~\ref{fig4}(b).  
	 
\subsection{The $ x x $ interaction}
\label{subsec:xx interaction}
\begin{figure}[]
	\subfloat[]{
		\includegraphics[scale=0.5, width=0.5\linewidth]{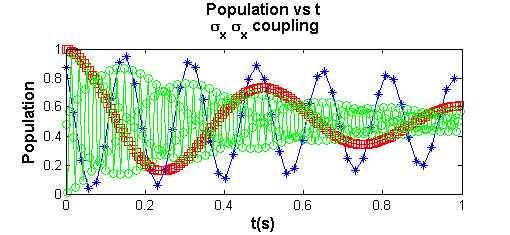}}
	\subfloat[ ]{
		\includegraphics[scale=1, width=0.5\linewidth]{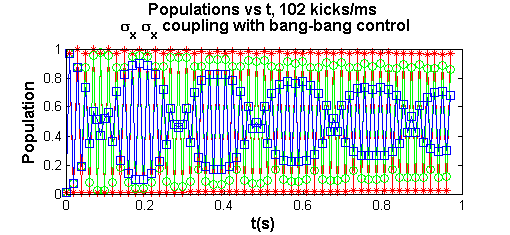}}
	\caption{Time evolution of the population level $ \ket{0}, \rho_{00} (t) $.  (a) at different kick rates. Stars indicate the case where $ \Gamma=52$ kicks/s,squares indicate $ \Gamma=152$ kicks/s  and circles indicate $ \Gamma=202$ kicks/s. (b) at different decoupling frequencies.Stars indicate the case where this frequency is 25.5 Hz ,circles indicate 10.2 Hz  and squares indicate 5.10 Hz. Kick rate $ \Gamma$=102 kicks/s. Again, $ \Omega/2$ is taken to be  150  Hz in both cases. } 
	\label{fig5}		
\end{figure}

For the $ x x $ interaction, we work in the $ \{\ket{+},\ket{-}\} $ basis. Hence, we start from the initial state $ \rho^S(0)_{\{\ket{+},\ket{-}\}}=\frac{1}{2}(I+\sigma_x) $ which is equivalent to $ \rho^S(0)_{\{\ket{0},\ket{1}\}}=\frac{1}{2}(I+\sigma_z) $ in the computational basis. The environmental qubit is  assumed to be in the thermal equilibrium state $ \rho^E(0)=\frac{1}{2}(I+\sigma_z) $. In this case, we monitor the populations of the two states $ \ket{0}$ and $\ket{1}\} $ reflected in the diagonal entries $ \rho^S_{00},\rho^S_{11} $ and look for the oscillations predicted by equation (\ref{oscill}). We observe that for higher kick rates, the oscillations damp quickly and approach the value 0.5, as predicted. This is illustrated in figure ~\ref{fig5}(a). Figure ~\ref{fig5}(b) depicts the sustained oscillations on applying a dynamical decoupling sequence to the system qubit. We also note that the frequency of oscillation decreases as one approaches the resonant kick rate $ \Omega/2 $ and then increases again.

\begin{figure}[]
	\subfloat[]{
		\includegraphics[width=0.5\linewidth]{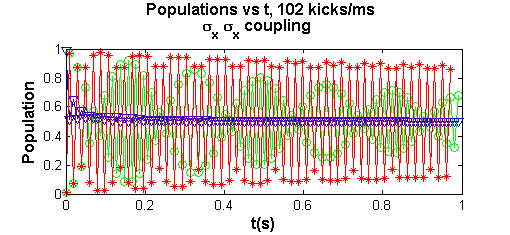}}
	\subfloat[]{\includegraphics[width=0.5\linewidth]{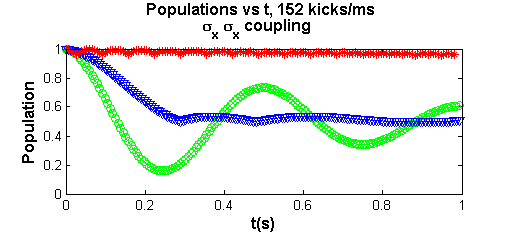}}
	
	\caption{Time evolution of the population level $ \ket{0}, \rho_{00} (t) $ (a) with $ \Gamma $=102 kicks/s. Stars indicate the case periodic where $ \pi$ pulses are applied at 10.2 Hz, circles indicate only Cory's kicks, and triangles indicate Cory's kicks and a random $ \pi $ pulse sequence. (b) with $ \Gamma\approx \Omega/2 \equiv$152 kicks/s. Stars indicate the case where periodic $ \pi$ pulses are applied at 15.2 Hz, circles indicate only Cory's kicks, and triangles indicate Cory's kicks and a random $ \pi $ pulse sequence. } 
	\label{fig6}		
\end{figure}

In this situation, applying a random  $ \pi $ pulse sequence causes the oscillations in the system population to damp very quickly to it's equilibrium value $ \frac{1}{2} $(shown in figure ~\ref{fig6}(a)). However, on applying a DD sequence when the kick rate is close to the resonant value, we observe that the system's initial population is conserved, with only a slight leakage over time(figure ~\ref{fig6}(b)). This effect can be explained by considering how the DD sequence works. If the DD sequence is applied at a time $ t $, it attempts to keep the system from evolving from the state in which it existed at time $ t $. Hence, the point of time at which the DD sequence starts is of crucial importance. In other words, the correlation time of the $ \pi $ pulse sequence must be much shorter than the correlation time of the noise to be suppressed. As mentioned before, at the resonant values of the kick rates, the frequency of oscillations decrease. Hence, when the first $ \pi $ pulse of the DD sequence is applied, the system is still not very far away from it's initial state. Thus, the DD sequence is able to arrest the evolution of the system very close to it's initial state.

\section{Conclusion}
\label{sec:conclusion}
As opposed to existing models for phase damping, a model using $ x x $ coupling was proposed which can successfully model amplitude damping. This, along with techniques developed by Cory and  Kondo, facilitates a controlled study of decoherence using finite resources. It is important to study both kinds of decoherence, phase and amplitude damping, in order to facilitate the development of control strategies to suppress a general decoherence process.
 
We have also shown that on combining the random kick model and the randomized $ \pi $ pulse sequence method, we obtain faster decoherence rates in the phase damping case. However, when the kick rate is close to the system-environment coupling, it is observed that decoherence is suppressed on applying a temporally randomized $ \pi $ pulse sequence. Viola and Knill \cite{viola} have previously identified cases where a random dynamical decoupling sequences can have more relaxed time scale requirements as compared to periodic $ \pi $ pulses and can become superior to existing techniques. The results obtained here demonstrate that a random $ \pi $ pulse sequence to the environmental qubit can help in the preservation of system coherence, sometimes even outperforming a periodic $ \pi $ pulse sequence. In the case of $ x x $ interaction, the combination of the two randomizing techniques achieve a faster decoherence rate. Due to decreased oscillation frequency when the kick rate is close to the coupling to the environment, dynamical decoupling proves to be very effective at this kick rate. 
\section*{Acknowledgements}
I thank Dr TS Mahesh for discussions and his valuable comments. 

\begin{thebibliography}{10}
	\bibitem{qerr}
	Peter W. Shor, Phys. Rev. A 52, R 2493(R) (1995) 	
	\bibitem{dd}
	Lorenza Viola and Seth Lloyd
	Phys. Rev. A 58, 2733 (1998)	
	\bibitem{jones} Jonathan A. Jones, Progress in Nuclear Magnetic Resonance Spectroscopy Volume 59, Issue 2, August 2011, Pages 91 - 120
	\bibitem{zurek}
	W. H. Zurek, Phys. Rev. D 26, 1862(1982)
	\bibitem{Cory}
	G. Teklemariam, E. Fortunato, C. L´ opez, J. Emerson,
	J. P. Paz, T. Havel, and D. Cory, Phys. Rev. A 67, 062316 (2003).
	\bibitem{kondo}
	Yasushi Kondo, Mikio Nakahara, Shogo Tanimura, Sachiko Kitajima,Chikako Uchiyama, and Fumiaki Shibata, J. Phys. Soc. Jpn. 76, 074002 (2007)
	\bibitem{swathi}
	Hegde, Swathi S, Mahesh, T. S., Phys. Rev. A .89.062317(2014)
	
	\bibitem{viola}
	Lorenza Viola,Emanuel Knill, Phys. Rev. Lett., 94,060502 (2005)
	\bibitem{nielson}
	M. A. Nielsen and I. L. Chuang, Quantum Computation and Quantum Information (Cambridge
	University Press, Cambridge, 2000)
\end{thebibliography}

\end{document}